\documentclass[english,aps,prapplied,twocolumn,superscriptaddress,longbibliography]{revtex4-2}
\usepackage[urlcolor=blue, hyperindex, colorlinks, bookmarks=true]{hyperref}
\usepackage{graphicx}
\usepackage{times}
\usepackage{amsmath}
\usepackage{braket}
\usepackage{xcolor}
\usepackage{tabularx}
\usepackage[symbol]{footmisc}

\usepackage{bm}
\usepackage{bbm}
\usepackage{natbib}
\usepackage{url}
\usepackage{textcase}

\definecolor{new}{rgb}{.38,.6,.38}
\definecolor{old}{rgb}{1,0,0}

\begin{document}

\title{Spin digitizer for high-fidelity readout of a cavity-coupled silicon triple quantum dot}

\author{F. Borjans}
\affiliation{Department of Physics, Princeton University, Princeton, New Jersey 08544, USA}
\author{X. Mi}
\altaffiliation{Present address: Google Inc., Santa Barbara, California 93117, USA}
\affiliation{Department of Physics, Princeton University, Princeton, New Jersey 08544, USA}
\author{J. R. Petta}
\affiliation{Department of Physics, Princeton University, Princeton, New Jersey 08544, USA}
\email{petta@princeton.edu}
\date{\today}

\begin{abstract}
{
An important requirement for spin-based quantum information processing is reliable and fast readout of electron spin states, allowing for feedback and error correction. However, common readout techniques often require additional gate structures hindering device scaling or impose stringent constraints on the tuning configuration of the sensed quantum dots. Here, we operate an in-line charge sensor within a triple quantum dot, where one of the dots is coupled to a microwave cavity and used to readout the charge states of the other two dots. Owing to the proximity of the charge sensor, we observe a near-digital sensor response with a power signal-to-noise ratio $>$450 at an integration time of $t_{\rm int}$ = 1 $\mu$s. Despite small singlet-triplet splittings $\approx$40 $\mu$eV, we further utilize the sensor to measure the spin relaxation time of a singlet-triplet qubit, achieving an average single-shot spin readout fidelity $>$99\%. Our approach enables high-fidelity spin readout, combining minimal device overhead with flexible qubit operation in semiconductor quantum devices.
}
\end{abstract}

\maketitle

\section{Introduction}

High-fidelity control of single spin qubits is becoming the norm in silicon quantum dots \cite{yoneda_quantum-dot_2018,yang_silicon_2019,andrews_quantifying_2019,huang_fidelity_2019,xue_benchmarking_2019}. In addition to the precise control of electron spin states, fast high-fidelity readout is a strict requirement to realize logical qubit codes by extending coherence through dynamic feedback and error-correction \cite{fowler_surface_2012}. Electron spin readout is often performed through spin-to-charge conversion with dedicated charge sensors that are probed with dc currents \cite{elzerman_single-shot_2004,morello_single-shot_2010}, ac modulated signals \cite{yang_dynamically_2011,jones_spin-blockade_2019}, or rf-reflectrometry \cite{reilly_fast_2007,Angus_gate-defined_2007,barthel_fast_2010,connors_rapid_2020}. Gate-based sensing approaches offer reduced device complexity, but often constrain device tuning to maximize the sensor signal \cite{colless_dispersive_2013,gonzalez-zalba_probing_2015,urdampilleta_gate-based_2019,west_gate-based_2019,zheng_rapid_2019}.

In recent years, a powerful hybrid circuit quantum electrodynamics architecture has emerged, integrating spin qubits with on-chip microwave cavities \cite{petersson_circuit_2012}. Recently, strong spin-photon coupling \cite{mi_coherent_2018,samkharadze_strong_2018,landig_coherent_2018} and resonant coupling between distant spins has been observed \cite{borjans_resonant_2020}. Moreover, dispersive interactions have been utilized to readout spin states by probing the cavity transmission amplitude and phase \cite{petersson_circuit_2012,mi_coherent_2018,zheng_rapid_2019}. However, even these approaches only perform well with specific tuning of the quantum dots, impeding general qubit flexibility, especially when additionally constrained by valley splitting.

In this work, we study a triple quantum dot (TQD) coupled to a cavity via a split-gate cavity coupler \cite{borjans_split-gate_2020}. By placing one quantum dot chemical potential on resonance with its neighboring lead, the cavity voltage excites resonant tunneling events, loading the cavity and reducing its quality factor and transmitted amplitude. We use the capacitive coupling between the quantum dots in the array to sensitively probe the charge occupation in the remaining two quantum dots and find large power signal-to-noise ratios at integration times of only one microsecond. We further probe the two electron dynamics in the system via pulsed spectroscopy, extracting the singlet-triplet energy splitting in the array. Finally, we extract the triplet relaxation times $T_1$ allowing estimates for promising spin readout fidelities using this readout approach.

\begin{figure*}[htb]
\centering
\includegraphics[width=2\columnwidth]{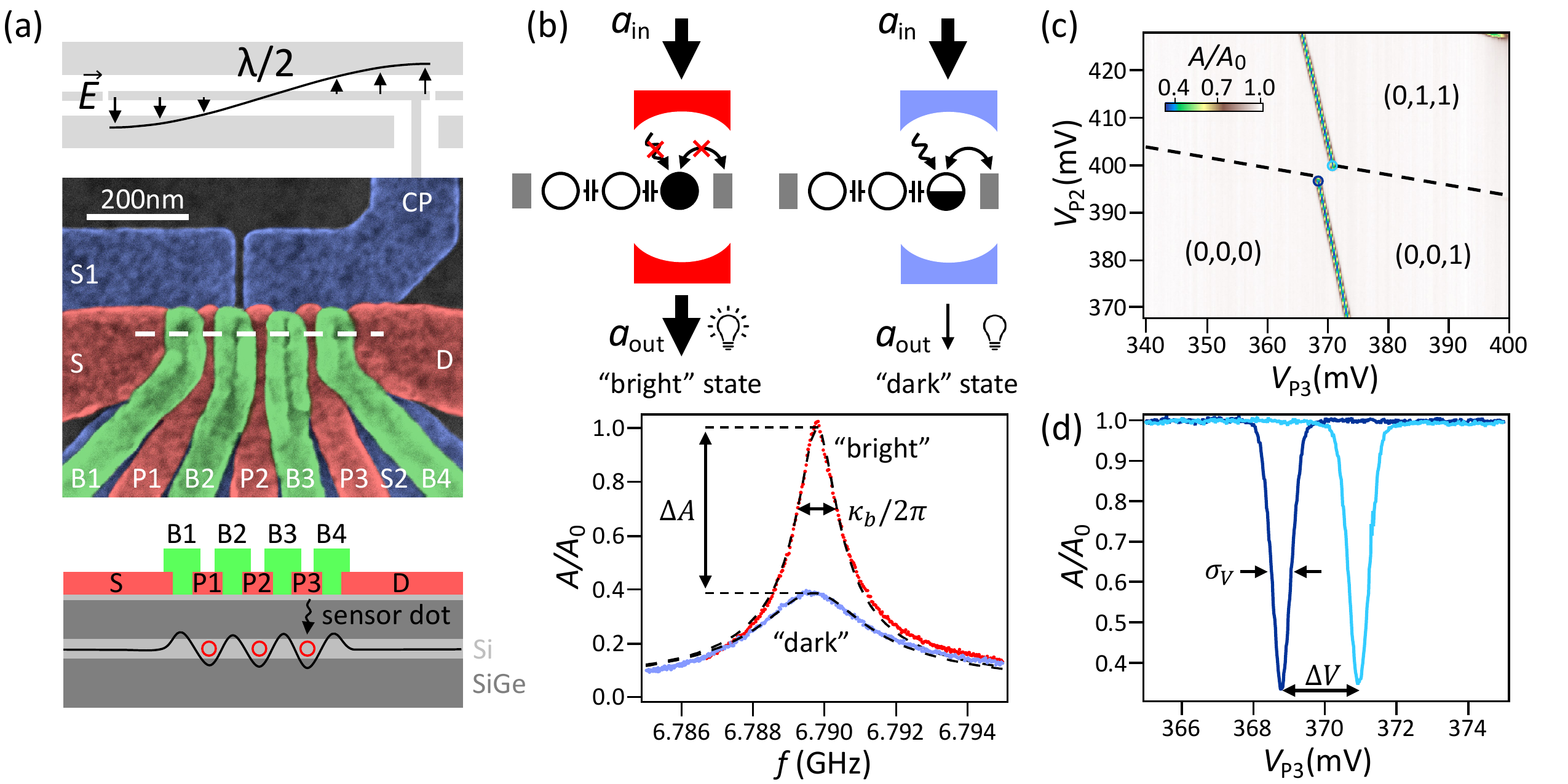}
\caption{Nearly-digital charge detection. (a) False-colored scanning electron micrograph of the TQD device. The electric field of the cavity schematically illustrated at the top couples to the device through the CP-gate. Three Al gate layers define the TQD confinement potential shown in the bottom cross-section through the active area of the device (white dashed line). (b) The cavity transmission amplitude $A/A_0$ is gated by the charge dynamics between dot 3 and the drain, D. In Coulomb blockade, the cavity and sensor dot are decoupled leading to a large $A/A_0$ (``bright" state). When tunneling between dot 3 and D is permitted ($\mu_3=\mu_{\rm D}$), the charge dynamics lead to dissipation and reduce $A/A_0$ (``dark" state). (c) $A/A_0$ as a function of $V_{\rm P2}$ and $V_{\rm P3}$, while dot 1 is empty. The uncoupled dot-lead transition for dot 2 is indicated by a dashed line. (d) $A/A_0$ as a function of $V_{\rm P3}$, at $V_{\rm P2}$ indicated by colored circles in (c). Loading an electron on dot 2 causes a capacitive shift of $\mu_3$, resulting in a large voltage shift $\Delta V=2.1\,\rm mV$ that can digitally gate the cavity transmission.}
\label{fig:1}
\end{figure*}

\section{Digital charge and spin sensing}
Our measurement approch couples the electric field of a half-wavelength $\lambda/2$ coplanar waveguide (CPW) cavity to the chemical potential of a sensor quantum dot via a coupling gate, CP \cite{borjans_resonant_2020}. The sensor quantum dot is situated inside a TQD device that is defined with three layers of overlapping aluminum gate electrodes on top of an isotopically enriched Si/SiGe wafer [Fig.\ \ref{fig:1}(a)]. We use the notation $(N_1,N_2,N_3)$ to indicate the number of electrons $N_i$ in each dot. The chemical potential of the sensor dot $\mu_3$ is tuned close to the Fermi level of a drain reservior, allowing an electron to be loaded and unloaded from the neighboring reservoir. When the sensor dot is in Coulomb blockade, it is decoupled from the cavity and the cavity transmission amplitude $A/A_0$ is large [Fig.\ \ref{fig:1}(b), red trace] \cite{petersson_circuit_2012}. We observe a resonance with $f_r=6.79\,\rm GHz$ and linewidth $\kappa_b/2\pi = 1.07\,\rm MHz$. We define this configuration as the ``bright" state of the system. When $\mu_3$ and the chemical potential of the drain reservoir $\mu_{\rm D}$ are aligned, cavity-field-induced tunneling events lead to energy dissipation and damp the cavity response. The reduced signal is shown in blue in Fig.\ \ref{fig:1}(b), with a relative signal strength of about 0.4 and $\kappa_d/2\pi=2.8\,\rm MHz$. Correspondingly, this configuration represents the ``dark" state of our system. The precise alignment of chemical potentials required to access this ``dark" state is well-suited for use as a sensitive charge sensor, as $\mu_3$ is strongly affected by capacitive interactions with adjacent dots in the TQD.

The digital nature of the charge readout is illustrated in Figs.\ \ref{fig:1}(c, d).  Figure \ref{fig:1}(c) shows  $A/A_0$ as a function of plunger gate voltages $V_{\rm P2}$ and $V_{\rm P3}$ with dot 1 unoccupied. Since dot 3 is directly coupled to the cavity through gate CP, the dot 3 charge transition causes a strong signal in $A/A_0$. We tune the tunneling rate of the dot 3-drain reservoir charge transition via barrier gate voltage $V_{\rm B4}$ to be resonant with the cavity frequency, minimizing $A/A_0\approx 0.4$ [see Appendix \ref{app:barriertuning}]. Due to the lack of a reservoir for dot 2 with a vacant dot 1, the dot 2 charge transition is not visible in $A/A_0$. However, the capacitive shift of $\mu_3$ caused by a change in $N_2$ is clearly visible. It is this large mutual capactiance that we harness to achieve high fidelity readout of dots 1 and 2, using dot 3 as the charge sensor. We acquire scans of $A/A_0$ as a function of $V_{\rm P3}$ in Fig.\ \ref{fig:1}(d) for the values of $V_{\rm P2}$ indicated by the colored circles in Fig.\ \ref{fig:1}(c). The dot 3 charge transition shifts by about $\Delta V=2.1\,\rm mV$ when an electron is added to dot 2, a shift that is much greater than the width of the charge transition $\sigma_V=0.27\,\rm mV$.

\section{Large scale charge state readout}

\begin{figure*}[t!]
\includegraphics[width=2\columnwidth]{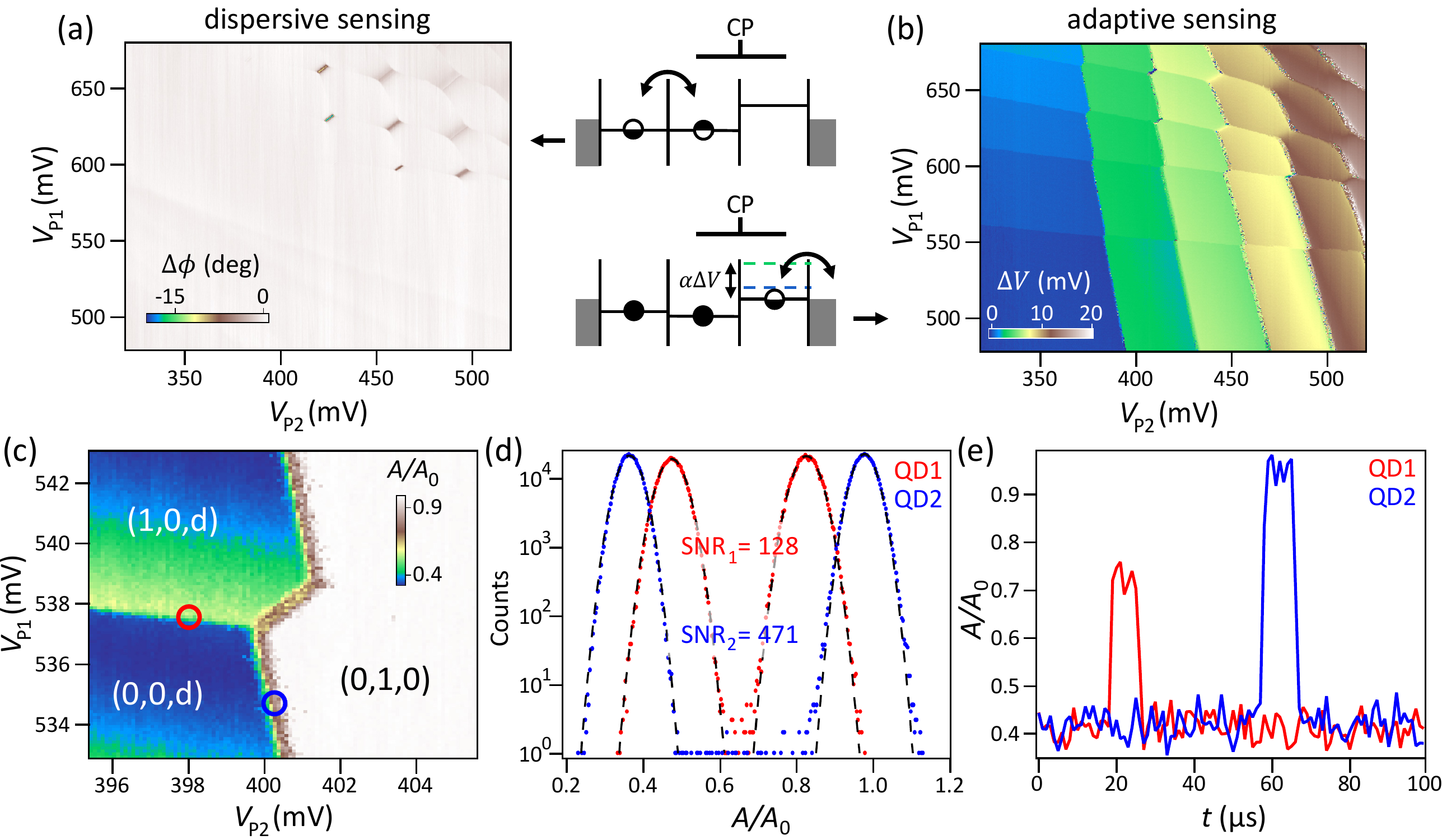}
\caption{Adaptive sensing of a DQD. (a) Dispersive phase response $\Delta\phi$ as a function of $V_{\rm P1}$ and $V_{\rm P2}$. Interdot transitions between dots 1 and 2 are visible in only a portion of the plot. (b) Charge sensing using dot 3 as a charge sensor and mapping out $\Delta V$ over the same range of $V_{\rm P1}$ and $V_{\rm P2}$ as in (a). Changes in $N_1$ and $N_2$ shift $\mu_3$ due to mutual capacitance, requiring a compensation $\alpha\Delta V$ with energy lever arm $\alpha$ to maintain the ``dark" state. We refer to this mode of operation as ``adaptive sensing." The blue and green dashed lines indicate qualitative $\mu_3$ shifts corresponding to the addition of an electron to dot 1 or 2. (c) $A/A_0$ measured as a function of $V_{\rm P1}$ and $V_{\rm P2}$ in the single electron regime without dynamic tracking. (d) Histograms of $A/A_0$ acquired at $1\,\rm MS/s$ for one second on the $(0,0,d)-(1,0,d)$ and $(0,0,d)-(0,1,0)$ transitions. We extract a power SNR = 128 for dot 1 and 471 for dot 2. (e) Time traces of corresponding electron hopping events of $<10~ \mu$s duration.}
\label{fig:2}
\end{figure*}

We now highlight the differences between conventional dispersive sensing and our digital readout approach. Figure \ref{fig:2}(a) shows the charge stability diagram of a double quantum dot (DQD) formed from dots 1 and 2 using a conventional dispersive sensing approach. We measure the cavity phase shift $\Delta\phi$ as a function of $V_{\rm P1}$ and $V_{\rm P2}$. As with cavity measurements of other systems, it is difficult to directly resolve charge transitions that change the total number of electrons in the DQD \cite{frey_dipole_2012}. The interdot charge transitions are sensed by the coherent coupling to the cavity photons, shifting the resonator frequency and causing a maximal phase shift when the transition energy is nearly resonant with the cavity, i.e. $2t_{c,12}\approx hf_r$, where $t_{c,12}$ is the tunnel coupling between dots 1 and 2, and $h$ is Planck's constant. An absence of interdot transitions can be caused either by a lack of charge transitions (i.e.\ the DQD is empty) or a large detuning between the charge transition energies and energy of a cavity photon.

Utilizing the capacitive interaction between the DQD and sensor dot, we now measure the charge stability diagram of the DQD by mapping out the voltage shift $\Delta V$ for each point in the same ($V_{\rm P1}$,$V_{\rm P2}$) gate voltage space [Fig.\ \ref{fig:2}(b)] \cite{chanrion_charge_2020}. We do so by repeating measurements similar to those shown in Fig.~\ref{fig:1}(d) for each point gate voltage space, tracking the sensor dot charge transition throughout the scan. To compensate for the capacitive coupling of $V_{\rm P1}$ and $V_{\rm P2}$ to $\mu_3$, we subtract a corresponding plane from the data. We define $\Delta V=0$ in the bottom left corner, where we find the $(0,0,d)$ charge state. We denote the resonant dark state $\mu_3=\mu_{\rm D}$ with $N_3$ fluctuating between $0$ and $1$ as $d$. A comparison of Figs.~\ref{fig:2}(a) and (b) highlights the advantages of this sensing approach: In contrast to the dispersive sensing method, we clearly see dot-lead charge transitions that change total electron number. Moreover, we observe a number of interdot charge transitions in Fig.\ \ref{fig:2}(b) that are not visible in Fig.\ \ref{fig:2}(a) due to the mismatch between interdot tunnel coupling and the photon energy.

\section{Performance of the charge sensor}

We now characterize the signal-to-noise ratio (SNR) of the charge sensor. To permit high-bandwidth measurements, we fix $V_{\rm P3}$ in Fig.\ \ref{fig:2}(c) such that it is resonant in the $(0,0,d)$ charge regime and compensate for the linear coupling to $V_{\rm P2}$. When the DQD charge occupation changes, $\mu_3$ shifts and we see a change in $A/A_0$. As expected, the signal is weaker for the $(0,0,d)-(1,0,d)$ charge transition relative to the $(0,0,d)-(0,1,0)$ transition due the extra distance between dot 1 and the sensor. The sensing signal on the $(0,0,d)-(0,1,0)$ charge transition is extremely robust, i.e.\ the signal shifts from fully resonant ($A/A_0 \approx 0.4$,``dark") to complete Coulomb blockade ($A/A_0 \approx 1$,``bright") [consistent with Fig.\ \ref{fig:1}(d)].

The power SNR for dot 1(2) is defined by
\begin{equation}
\mathrm{SNR_{1(2)}} = \left(\frac{\Delta A_{1(2)}}{\sigma_{A,1(2)}}\right)^2,
\end{equation}
where $\Delta A_{1(2)}$ is the signal strength of the corresponding dot charge transition and $\sigma_{A,1(2)}$ is the r.m.s amplitude of the measurement noise at the sensing point \cite{stehlik_fast_2015}. We acquire a time series at the charge transition for one second with a $1\,\rm MS/s$ sampling rate and histogram $A/A_0$ in Fig.\ \ref{fig:2}(d) for the $(0,0,d)-(1,0,d)$ transition (red) and the $(0,0,d)-(0,1,0)$ transition (blue). We analyze the data by fitting a double-Gaussian distribution with standard deviation $\sigma_A$ for both peaks. For dot 1, we find $\Delta A_1=0.35$ and $\sigma_{A,1}=0.031$, yielding $\mathrm{SNR_1}=128$. Similarly, for dot 2 we find $\Delta A_2=0.61$, $\sigma_{A,2}=0.028$ and $\mathrm{SNR_2}=471$. These values compare favorably to previous on-chip resonator experiments at the same integration time ($\mathrm{SNR}=6$ \cite{zheng_rapid_2019} and $\mathrm{SNR}=190$ \cite{stehlik_fast_2015}) and state-of-the-art reflectometry setups using an external sensor ($\mathrm{SNR}=184$ \cite{connors_rapid_2020}, estimated assuming white noise). The SNR using the unfiltered signal corresponds to a bandwidth of $f_{\rm SW}=500\,\rm kHz$ due to the Nyquist-theorem. Depending on the measurement bandwidth $f_{\rm BW}$, the integrated noise amplitude changes according to
\begin{equation}
\sigma_{A,1(2)}^2(f_{\rm BW}) = \int\limits_{1\,\rm Hz}^{f_{\rm BW}}\left[S_{A,1(2)}(f)\right]^2\mathrm{d}f,
\end{equation}
with the amplitude spectral density $S_{A,1(2)}(f)$. We calculate $\sigma_{A,1(2)}^2(f_{\rm BW})$ using noise spectra acquired on the side of the dot 3 resonance ($A/A_0\approx 0.7$) and on resonance (at $A/A_0\approx 0.4$), obtaining $\sigma_{A,1(2)}=0.034(0.027)$, which is in good agreement with the fit parameters from Fig.\ \ref{fig:2}(d). We estimate the bandwidth at which $\mathrm{SNR}_{1(2)}=1$ by extending the noise spectra with matching white noise until $\sigma_{A,1(2)}=\Delta A_{1(2)}$. We obtain $\mathrm{BW_1}=89\,\rm MHz$ and $\mathrm{BW_2}=271\,\rm MHz$, corresponding to minimum integration times $t_{\rm min,1}=5.6\,\rm ns$ and $t_{\rm min,2}=1.8\,\rm ns$.
We note that these values are theoretical, as the physically attainable bandwidth is in practice limited by the bandwidth of the cavity, $\kappa_b/2\pi$. Two time traces of charge sensing are plotted in Fig.\ \ref{fig:2}(e), demonstrating that the full bandwidth of the circuit can be used to measure a clear charge signal.

\section{Single shot spin readout}

\begin{figure}[t!]
\includegraphics[width=\columnwidth]{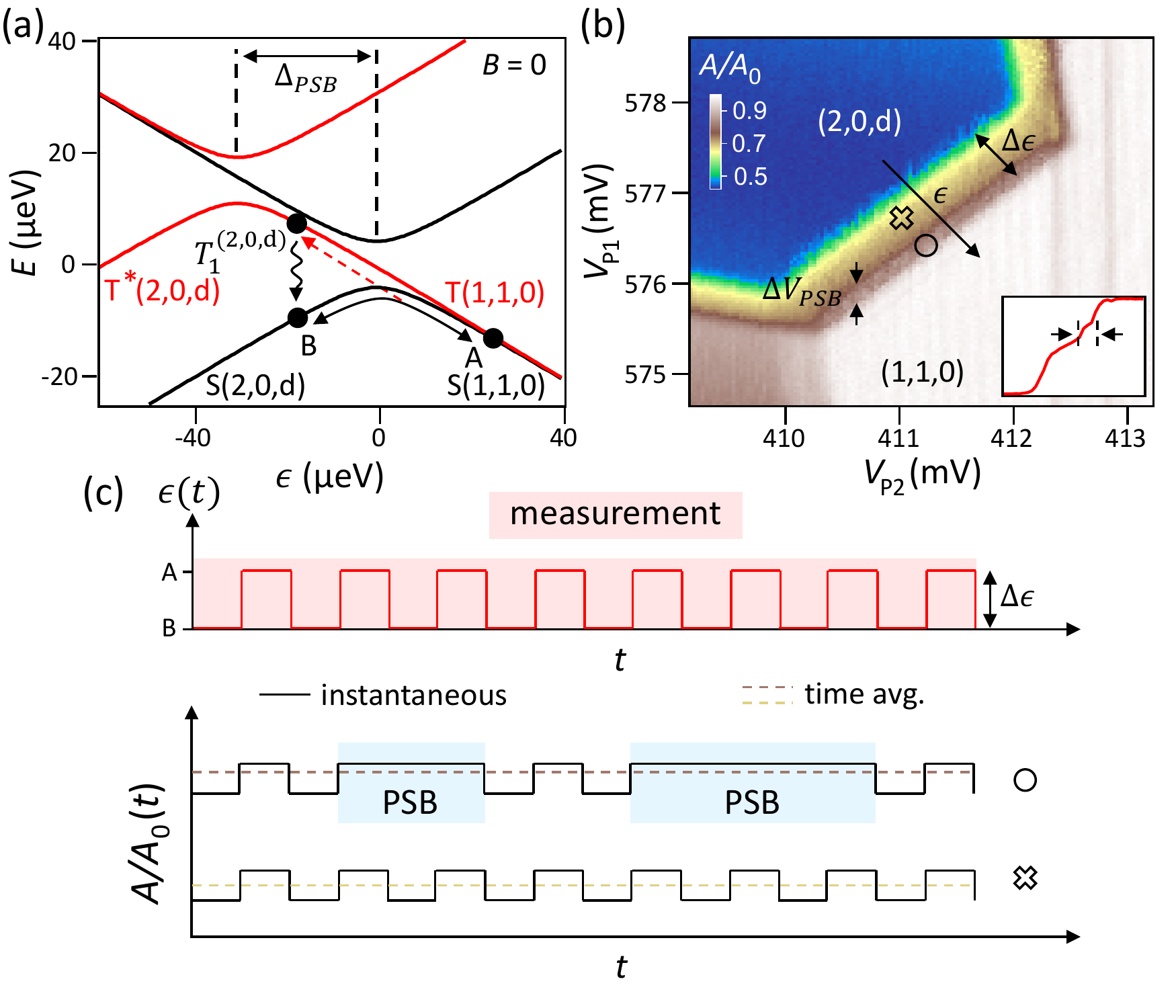}
\caption{Pauli spin blockade in a singlet-triplet qubit. (a) Singlet-triplet energy level diagram near the $(2,0,d)-(1,1,0)$ interdot charge transition with $B$ = 0. For $-\Delta_{\rm PSB} < \epsilon < 0$, only singlets access the $(2,0,d)$ charge ground state. (b) Averaged signal when a continuous rectangular pulse described in (c) is applied across the interdot detuning axis. The brown region near the $(1,1,0)$ charge state is due to PSB. A linecut through the data along $\epsilon$ (inset) shows the PSB signal as an additional step indicated by the arrows. (c) A continuous train of square pulses is applied to $\epsilon$, inducing charge transitions dependent on the starting position in gate voltage space. Data are signal averages over multiple complete cycles (shaded light red). Four different zones of charge dynamics are observed. Deep in $(2,0,d)$ and $(1,1,0)$ the charge state and corresponding signal is static ``dark" (``bright"). For $\epsilon<-\Delta_{\rm PSB}$ (cross) both spin states can access the $(2,0,d)$ charge state and charge hops are always observed when pulsing, while for $-\Delta_{\rm PSB} < \epsilon < 0$ (circle) the charge dynamics are spin dependent, causing a shift in the time-averaged signal.}
\label{fig:3}
\end{figure}

Fast charge readout is especially powerful when used in combination with spin-to-charge conversion to allow for fast electron spin state readout. A scalable approach is the use of Pauli spin blockade (PSB) to distinguish two-electron singlet and triplet states in a DQD \cite{ono_current_2002,petta_coherent_2005}. Figure \ref{fig:3}(a) shows the energies of the two-electron spin states as a function of level detuning $\epsilon$. As the experiment is conducted at zero magnetic field $(B=0)$, the triplet states are degenerate, and the level diagram consists of a singlet state manifold (black) and a triplet state manifold (red). By applying a square pulse to $\epsilon$, we can stimulate spin-conserving tunneling events between the $(1,1,0)$ and $(2,0,d)$ states (see points A and B). If point A is chosen such that the pulse starts deep in the $(1,1,0)$ regime, random mixing of the separated electrons allows the initialization of a mixed state consisting of $S(1,1,0)$ and $T(1,1,0)$. With a pulsed detuning excursion to point point B, a singlet state will tunnel from $S(1,1,0)$ to $S(2,0,d)$, while the triplet will remain in $T(1,1,0)$ due to Pauli blockade, causing a visible difference in the charge signal \cite{johnson_tripletsinglet_2005}. Charge readout can therefore be used to differentiate spin singlet and triplet states.

We tune dots 1 and 2 near the $(1,1,0)-(2,0,d)$ transition and apply a continuous train of 0.5 mV square pulses with period $T=400\,\rm\mu s$ and 50\% duty cycle to the gates $(V_{\rm P1},V_{\rm P2})$ to realize the diagonal detuning pulses. We measure the average $A/A_0$ in Fig.\ \ref{fig:3}(b) as a function of $V_{\rm P1}$ and $V_{\rm P2}$ \cite{prance_single-shot_2012}. In the dynamic region between the $(1,1,0)$ and $(2,0,d)$ states, where the detuning pulse has an amplitude sufficient to cross the interdot charge transition, we observe a region in which the $(1,1,0)$ signal is enhanced (circle) relative to the neighboring region [cross, see Fig.\ \ref{fig:3}(c)]. The signal in this region indicates PSB, as the Pauli-blocked triplet states increase the fraction of time the system resides in $(1,1,0)$ [see A and B in Fig.\ \ref{fig:3}(a)]. The brown PSB region corresponds to an additional plateau in $A/A_0$ as shown by the linecut across the interdot transition in the inset of Fig.\ \ref{fig:3}(b).

\begin{figure}[t]
\includegraphics[width=\columnwidth]{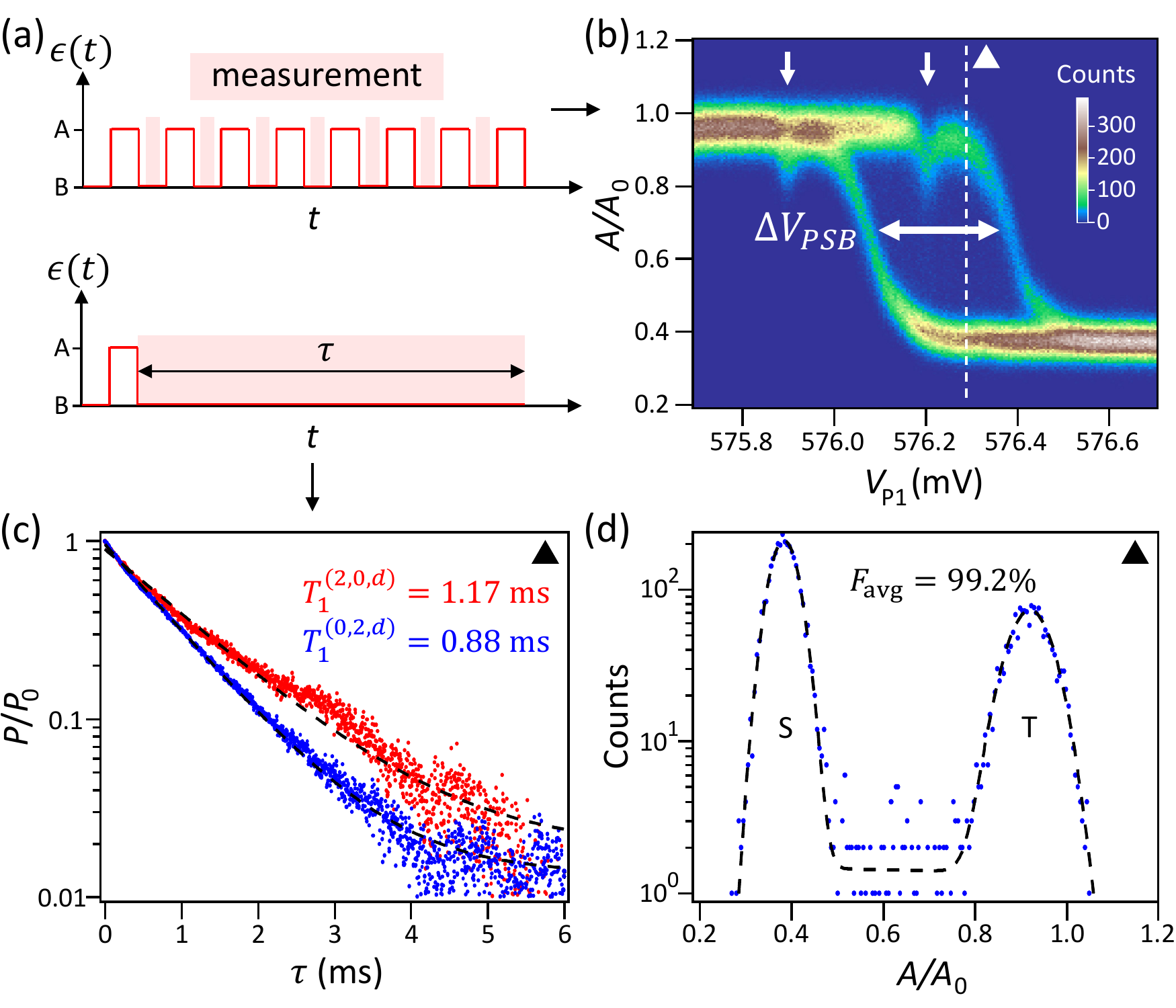}
\caption{Single shot readout using Pauli spin blockade. (a) Pulse sequences used to characterize PSB and extract the spin liftetime $T_1$. Measurements are performed during the shaded intervals of the pulse sequence.  (b) Single shot analysis of the PSB region. Averaged $A/A_0$ are histogrammed as a function of $V_{\rm P1}$, showing charge transitions corresponding to $S(1,1,0)$-$S(2,0,d)$ and $T(1,1,0)$-$T^*(2,0,d)$. (c) Normalized triplet probability $P/P_0$ as a function of $\tau$. Triplet relaxation is measured at the readout point indicated by the triangle in (b). The time spent at point B is increased and data are acquired as a function of delay time $\tau$, ensemble averaging over $10,000$ shots. (d) Readout fidelity analysis performed using a histogram of $A/A_0$ taken from (b) at the white dashed line. The SNR obtained from the histogram combined with $T_1$ from (c) yields $F_{\rm avg}=99.2\%$, modeled by the black dashed line.}
\label{fig:4}
\end{figure}

To give further evidence this region is due to PSB, we perform single shot readout on the time-domain signal by only analyzing the data during the negative segment of the detuning pulse, i.e.\ at point B [see Fig.\ \ref{fig:4}(a)]. We average $A/A_0$ for $100\,\rm\mu s$ and histogram the data on the vertical axis as a function of $V_{\rm P1}$ in Fig.\ \ref{fig:4}(b) \cite{jones_spin-blockade_2019}. Since $\epsilon\propto -V_{\rm P1}$, the corresponding points A and B are being moved through the diagram in Fig.\ \ref{fig:3}(a) from right to left. As $V_{\rm P1}$ is increased, the measured charge state at point B starts in the $(1,1,0)$ regime, until the singlet states are able to tunnel into the $(2,0,d)$ state at $V_{\rm P1}\approx 576.1\,\rm mV$, while the triplet states are blocked. Eventually, for $V_{\rm P1}\approx 576.4\,\rm mV$, the spin blockade is lifted due to the presence of the excited triplet state $T^*(2,0,d)$ and the triplet states also tunnel into the $(2,0,d)$ state. In addition to the main charge transition behavior, we observe two features indicated with white arrows, which we attribute to the dispersive coupling of the interdot transition to the cavity, with $2t_c \ll hf_r$.

The histogrammed data in Fig.\ \ref{fig:4}(b) resemble two Fermi functions that are separated by a voltage $\Delta V_{\rm PSB}$ \cite{diCarlo_differential_2004,jones_spin-blockade_2019}. 
We extract $\Delta V_{\rm PSB}$ by taking a linecut at $A/A_0=0.65$ and fitting a double-Gaussian function. We are able to resolve these energy scales as the electron temperature remains low during these measurements with $T_e=60\,\rm mK$, measured using the width of dot-lead Fermi functions.
This well visible separation allows us to extract $\Delta_{\rm PSB}^{\rm (2,0,d)}=41\,\rm\mu eV$. Similar measurements pulsing from $(1,1,0)$ to $(0,2,d)$ yield $\Delta_{\rm PSB}^{\rm (0,2,d)}=36\,\rm\mu eV$ (data not shown). These energies are comparable with valley splittings extracted from similar devices \cite{mi_high-resolution_2017}, suggesting that $\Delta_{\rm PSB}$ is limited by the valley splitting in this device.

As a final demonstration of the utility of this readout approach, we measure the spin relaxation time $T_1$ of the triplet states. We do so by setting $V_{\rm P1}=576.3\,\rm mV$, indicated by the white triangle in Fig.\ \ref{fig:4}(b), and extend the dwell time spent in point B to $6\,\rm ms$ to probe the $T_1$ process after preparing a mixed state with a short pulse to point A [see lower panel of Fig.\ \ref{fig:4}(b)]. We plot the average of $N$ = 10,000 shots in Fig.~\ref{fig:4}(c) and renormalize the signal compared to the initial triplet probability $P_0$. We fit an exponential decay curve and extract a triplet relaxation time $T_1^{(2,0,d)}=1.17\,\rm ms$. Similarly, we extract $T_1^{(0,2,d)}=0.88\,\rm ms$ at the $(1,1,0)-(0,2,d)$ transition.

We combine these results to estimate the readout visibility. By analyzing the $A/A_0$-histogram at the readout point for the $T_1^{(2,0,d)}$ measurement [Fig.\ \ref{fig:4}(d)] we find $\Delta A=0.536$ and $\sigma_S=0.03$ and $\sigma_T=0.047$ for the charge signal corresponding to the singlet and triplet states. Taking into account the integration time $t_{\rm int}=100\,\rm \mu s$ and the measured $T_1=1.17\,\rm ms$, we determine our visibility $V$ and maximum average readout fidelity $F_{\rm avg}$ as described in Refs. \cite{west_gate-based_2019,zheng_rapid_2019} to be $V=98.4\%$ and $F_{\rm avg}=99.2\%$.

One reason for the high fidelity readout is the ideal placement of the sensor. Even though the total charge is conserved during spin readout, the full polarization of the dots towards the sensor maximizes the differential charge signal. In addition, as $\Delta V\gg \sigma_V$ for the interdot transition, the strength of the S-T signal is the full digital cavity signal of this architecture, not the difference between the dot 1 and dot 2 signals. Finally, we note that spin readout is performed with a cavity probe power of $20\,\rm dB$ less than that used during charge state readout. As the device design is optimized for strong coupling to the interdot transitions of dots 1 and 2, we observe features likely related to photon assisted tunneling at higher probe powers.

In future devices, a change of the cavity-coupler geometry with minimized direct coupling to dots 1 and 2 could allow for higher probe powers. This configuration could be achieved by shifting the gap between gates S1 and CP towards the right, limiting the coupling of CP to only dot 3. Combining $\Delta A_2$ and $\sigma_{A,2}$ from the charge sensing measurements (that were performed at higher powers) with $T_1^{(2,0,d)}=1.17\,\rm ms$, we expect a potential spin readout visibility of $99.98\%$ and an average readout fidelity of $99.99\%$ with a 1 $\mu$s integration time. Increasing the designed cavity bandwidth via the corresponding output port coupling rate could additionally improve the SNR and maximum readout bandwidth. The current necessity of a sensor-reservoir could be overcome by replacing it with an additional dot and tracking of the corresponding interdot transition.

\section{Conclusion}
We have demonstrated high-fidelity charge and spin readout utilizing an in-line charge sensor located inside a cavity-coupled TQD. We find exceptional charge sensitivity, allowing large-scale mapping of the charge stability space in the sensed DQD. The sensor is optimally placed to directly translate this charge sensitivity to high-fidelity spin readout based on PSB. Here, the strong signal allows for detailed blockade spectroscopy, enabling the extraction of the quantum dot valley splittings. We find a spin readout visibility of $98.4\%$ and an average spin readout fidelity of $99.2\%$.

By using one of the three dots as an ancilla for charge sensing rather than realizing dispersive cavity sensing, no resonance conditions apply to the sensed pair of electrons, retaining maximum flexibility for qubit operation. The strong differential signal does not require a latching mechanism \cite{petersson_quantum_2010,studenikin_enhanced_2012,harvey-collard_high-fidelity_2018}, limiting the only tunable reservoir requirement to the sensor. Future device architectures optimized to suppress the dispersive coupling to interdot charge transitions could allow for larger cavity probe powers, increasing the readout visibility even further. The device architecture may be of special interest for encoded TQD exchange-only spin qubits \cite{andrews_quantifying_2019}, where the tunnel-coupled sensor dot could be both part of the qubit and the charge sensor, removing the overhead of an external sensor.


\begin{acknowledgements}
We thank Lisa Edge and Clayton Jackson at HRL Laboratories LLC for providing the heterostructure used in these experiments. Research sponsored by Army Research Office grant W911NF-15-1-0149 and the Gordon and Betty Moore Foundation's EPiQS Initiative through grant GBMF4535. Devices were fabricated in the Princeton University Quantum Device Nanofabrication Laboratory, which is managed by the department of physics.
\end{acknowledgements}

\appendix

\section{Optimizing the cavity response \label{app:barriertuning}}
\begin{figure}[t]
\includegraphics[width=\columnwidth]{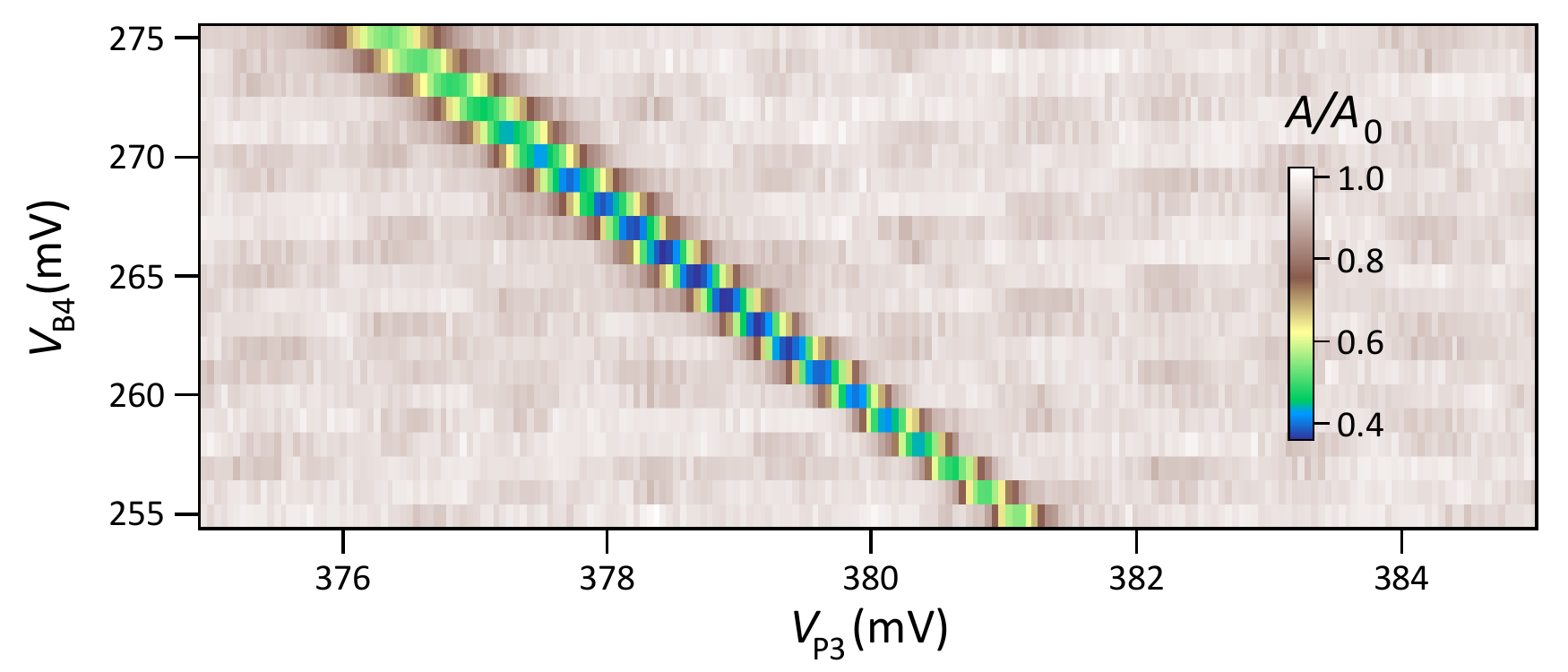}
\caption{Optimizing the cavity readout signal. The cavity signal $A/A_0$ is shown as a function of gate voltages $V_{\rm P3}$ and $V_{\rm B4}$. The visible reduction in microwave transmission corresponds to the condition $\mu_3=\mu_D$, where the first electron can both be loaded and unloaded from dot 3 via the drain reservoir. We observe a clear minimum of $A/A_0$ around $V_{\rm B4}\approx 265\,\rm mV$, away from which the signal strength decreases monotonically in both directions.}
\label{appfig:1}
\end{figure}

We optimize the cavity response in this experiment by tuning the barrier voltage such that the cavity transmission is minimal when the first electron transition is resonant with the drain reservoir ($\mu_3=\mu_D$) [see Fig.\ \ref{appfig:1}]. Besides the capacitive impact of the barrier gate voltage causing the transition to shift, we observe three zones in the data. For $V_{\rm B4}<260\,\rm mV$, the tunnel rate between dot 3 and the drain is too slow to efficiently couple to the cavity fields with frequency $f_r$. For $V_{\rm B4}>270\,\rm mV$ the large tunnel rates wash out the signal, again weakening the response. At about $V_{\rm B4}\approx 265\,\rm mV$ we observe a sweetspot between the two regimes, where the cavity coupling is most efficient and results in the best signal.

\bibliography{references}

\end{document}